\documentclass[pra,aps,twocolumn,groupedaddress,floats,showpacs,final,superscriptaddress]{revtex4-2}
\usepackage[latin3]{inputenc}
\usepackage[makeroom]{cancel}
\usepackage{graphicx}
\usepackage{amsmath}
\usepackage{amsfonts}
\usepackage{amssymb}
\usepackage{chngcntr}
\usepackage{color}
\usepackage[left=2cm,right=2cm,top=2cm,bottom=2cm]{geometry}
\usepackage{graphicx}
\usepackage{dcolumn}
\usepackage{bm}
\usepackage{simplewick}
\usepackage{array}
\usepackage{appendix}

\RequirePackage[
   hyperindex,colorlinks,bookmarksnumbered,
   plainpages=true,pdfstartview=FitH]{hyperref}
\hypersetup{linkcolor=blue,urlcolor=blue,citecolor=blue}
\usepackage{hyperref}

\definecolor{purple}{rgb}{0.5,0,0.6}

\begin{document}


\title{Role of Coulomb interaction in the valley photogalvanic effect}


\author{V.~M.~Kovalev}
\affiliation{Novosibirsk State Technical University, Novosibirsk 630073, Russia}

\author{A.~V.~Parafilo}
\affiliation{Center for Theoretical Physics of Complex Systems, Institute for Basic Science (IBS), Daejeon 34126, Korea}

\author{O.~V.~Kibis}
\affiliation{Novosibirsk State Technical University, Novosibirsk 630073, Russia}

\author{I.~G.~Savenko}
\affiliation{Department of Physics, Guangdong Technion--Israel Institute of Technology, 241 Daxue Road, Shantou, Guangdong 515063, China}
\affiliation{Technion -- Israel Institute of Technology, 32000 Haifa, Israel}
\affiliation{Guangdong Provincial Key Laboratory of Materials and Technologies for Energy Conversion, Guangdong Technion--Israel Institute of Technology, Guangdong 515063, China}


\begin{abstract}
We develop a theory of Coulomb interaction-related contribution to the photogalvanic current of the carriers of charge in two-dimensional non-centrosymmetric Dirac materials possessing a nontrivial structure of valleys and exposed to an external electromagnetic field.
The valley photogalvanic effect occurs here due to the trigonal warping of electrons and holes' dispersions in a given valley of the monolayer.
We study the low-frequency limit of the external field: The field frequency is smaller than the temperature $T$, and the electron-electron and electron-hole scattering times are much larger than the electron-impurity and hole-impurity scattering times.
In this regime, we employ the Boltzmann transport equations and show that electron-hole scattering dominates electron-electron scattering in intrinsic semiconductors.
A Coulomb electron-hole interaction-related contribution to the valley photogalvanic current can reduce the value of the bare photogalvanic current as electron and hole currents flow in opposite directions.
\end{abstract}

\maketitle


\section{Introduction}


The influence of the Coulomb scattering, namely, electron-electron (e-e) and electron-hole (e-h) interaction, on the transport properties of solids is an active research area.
At low temperatures, when particle-phonon scattering is frozen, the particle-particle scattering mechanisms may determine the temperate behavior of transport coefficients, particularly the Drude conductivity~\cite{Lithuanian}.
In Galilean-invariant systems with the parabolic spectrum, particle collisions do not affect the conductivity since the velocity is proportional to the particle momentum, and thus, the conservation of total momentum under e-e and h-h collisions results in the conservation of velocity.
Nevertheless, in strongly disordered samples, elastic scattering on impurities plays a sufficient role: impurities break the Galilean invariance,
and then, the influence of particle collisions not only can be finite but also results in strong Coulomb-induced corrections to conductivity~\cite{AltshulerAronovSpivak}.
In particular, weak localization corrections and corrections to magneto-oscillation transport phenomena~\cite{BEENAKKER19911}, including the Shubnikov--de Haas effect, can emerge.

\begin{figure*}[t!]
\includegraphics[width=1.8\columnwidth]{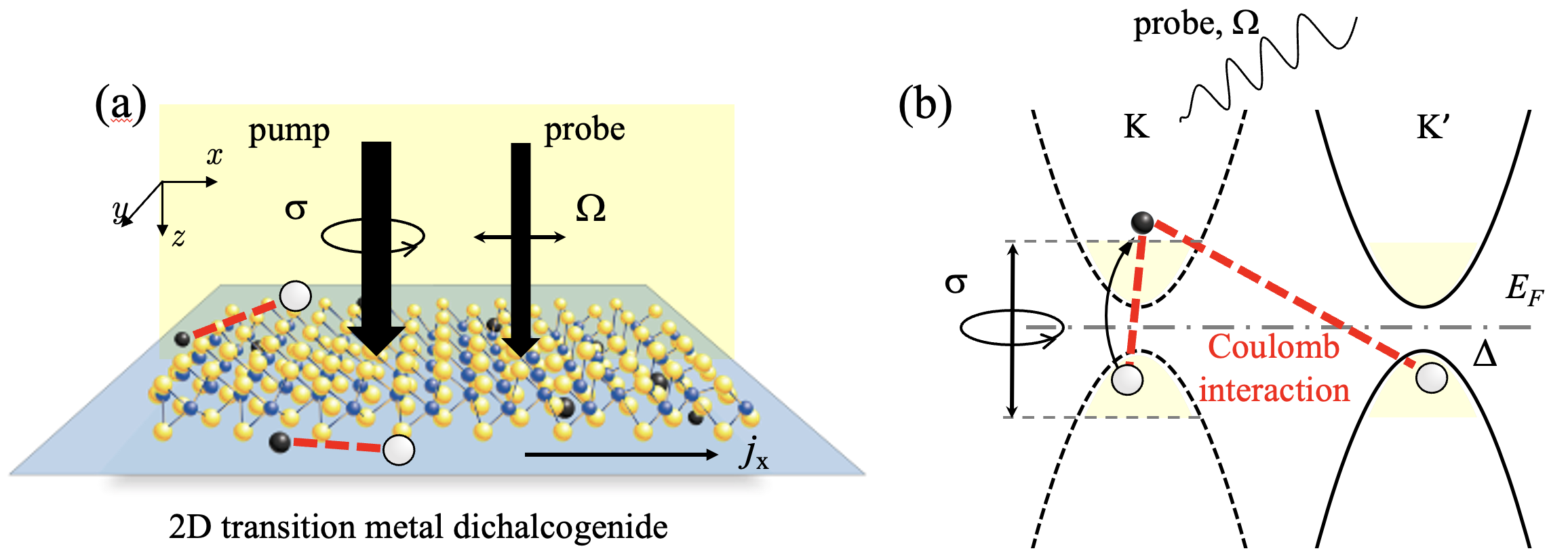}
\caption{(a) System schematic: a 2D material exposed to a linearly polarized electromagnetic (EM) field with the frequency $\Omega$ and a circularly polarized light with the polarization $\sigma$. A circular-polarized EM field changes electron population at a given valley due to photoinduced interband transition, whereas a linearly polarized EM results in a valley photogalvanic effect (vPGE).
(b) Reciprocal space of the 2D system, reflecting the multi-valley structure. Electrons scatter by holes in different valleys.}
\label{Fig1}
\end{figure*}

Interaction between charged particles can be repulsive or attractive, revealing different behavior in different temperature ranges.
Indeed, in addition to the direct Coulomb repulsion between identical particles (e-e or h-h) and attraction of different particles (e-h in semiconductors), the other interaction channels may also strongly affect the conductivity of the material.
In particular, strong particle-particle attraction in the Cooper channel leads to a renormalization of the Drude conductivity by superconducting fluctuations at sufficiently low temperatures (in the vicinity of the superconducting transition temperature).
This constitutes the phenomenon referred to as the paraconductivity~\cite{Varlamov}.
Furthermore, recent technological achievements open the way to the creation of ultra-clean nanostructures, where the electron (and hole) mean free path, which is usually limited by the interaction with impurities, is comparable with or even exceeds the sample width~\cite{superballistic_graphene,PhysRevB.106.L241302}.
The system is in the hydrodynamic regime of electron transport~\cite{GurzhiJETP, GurzhiUFN, Narozhny_review}.
In this regime, the particle momentum predominantly changes due to electron scattering off the sample boundaries.
In some materials, the e-e and e-h interaction starts to play a key role in particle transport, determining the viscosity of electron liquid.

This subject is especially timely for novel two-dimensional (2D) materials such as monolayers of graphene~\cite{RefGraphene2005, PhysRevLett.99.236809} and transition-metal dichalcogenides (TMD)~\cite{Mak1489, XiaoVHE, PhysRevB.77.235406}
--- the materials, which we consider.
A key specific property of these materials is the two-valley structure of the dispersion of carriers of charge.
Characteristic band structure results in the emergence of specific valley transport phenomena, such as the valley acoustoelectric effect and valley Hall effect~\cite{PhysRevLett.122.256801, PhysRevB.103.035434, PhysRevB.100.121405, PhysRevB.102.235405,
PhysRevB.102.155302, Kovalev2018NJP,PhysRevLett.132.096302}.
Recently, both the diffusive and hydrodynamic regimes of valley Hall and anomalous Hall effects have been intensively studied theoretically~\cite{Glazov2DMat22, PhysRevB.109.085301, arXiv:2310.17738}.
Photo-excitation of valley-dependent currents is possible even in conventional semiconductor devices such as silicon-based transistors~\cite{PhysRevB.83.121312, CitePure, Ivchenko2008}.
One of the critical properties of electrons and holes in the valleys of TMDs or a gapped graphene is the trigonal warping of valleys reflecting the  symmetry of the crystal lattice.
It results in the modification of the particle energy: $\epsilon_\mathbf{p}\approx p^2/2m+ \eta w(p_x^3-3p_xp_y^2)$, where $\mathbf{p}$ is the momentum of an electron or a hole, $m$ is the effective mass, $\eta=\pm1$ is the valley index, and $w$ is the warping parameter.
Thanks to that, in addition to the valley Hall effect, these materials reveal fascinating nonlinear transport phenomena such as the valley photogalvanic effect (vPGE)~\cite{PhysRevB.99.075405, Entin2019, Entin2, PhysRevB.107.085415}.
From the mathematical point of view, the vPGE represents the second-order response of the stationary electric current density to the external alternating EM field with a frequency much smaller than the material bandgap.

Various contributions to vPGE might exist at small temperatures.
For instance, attractive e-e interaction (the Cooper channel) can influence the magnitude of the vPGE~\cite{PhysRevB.106.144502}.
Namely, the emergence of superconducting fluctuations strongly affects the vPGE in the vicinity of the superconducting transition temperature, obeying the law $(T-T_c)^{-3}$ when $T \rightarrow T_c$ from above~\cite{PhysRevB.106.144502}.
The goal of this paper is to examine the vPGE in gapped Dirac materials due to particle-particle interaction in the Coulomb channel, when either the particle density is low enough or the temperature is higher than the superconducting transition temperature, thus the superconducting fluctuations do not yet play important role.

In a previous work by some of the authors of this paper~\cite{Entin2019}, bare vPGE in a non-degenerate electron gas was studied.
Here, we will assume that the electron and hole gases are non-degenerate, satisfying the Boltzmann statistics, and apply the two-band model with the parabolic dispersion of electrons and holes with equal effective masses and accounting for the warping-related corrections to the parabolic dispersion.
The main advantages of this model are that (i) it captures all the main physical effects and (ii) it can also be applied to a gapped monolayer graphene.

In an intrinsic gapped Dirac material at a finite temperature, three channels of particle-particle interaction can exist: e-e, h-h, and e-h.
In contrast to the gapless graphene case~\cite{PhysRevB.84.195408}, our calculations show that e-e and h-h contributions do not play a role in vPGE (at least in the framework of the two-band model).
Thus, we focus on e-h scattering as a dominating source of particle-particle interaction correction to the vPGE.

The paper is organized as follows. In Sec. II, we describe the general formalism and present the final expressions for the e-h interaction induced corrections to the vPGE current. In Sec. III we discuss the results. The last two sections contain the conclusion and acknowledgements.


\section{General formalism}
We will consider the diffusive regime, assuming that the temperature is low enough, thus the particle-impurity collision rate exceeds the particle-particle one~\cite{Lithuanian, [{If the temperature is low enough, the particle-phonon absorption processes are suppressed, whereas the possible particle-phonon emission processes can be incorporated into general scattering rate describing both the particle-impurity and particle-phonon emission processes}]FTN1}.
The EM field falls normally to the 2D monolayer (Fig.~\ref{Fig1}).
We will assume that the frequency of the external electromagnetic field, producing the vPGE (the probe), does not exceed the temperature, which determines the mean kinetic energies of electrons and holes.
Such an approximation allows us to consider the effect of the particle-particle collisions via successive approximations.

The second-order transport phenomena in general and particularly the vPGE are usually sensitive to the polarization of the EM field and the
symmetry of the system under study, namely, the time-reversal symmetry and the spatial inversion symmetry. The phenomenological relation connecting the photoinduced rectified electric current and the amplitude of external probe EM field reads $j_\alpha=\lambda_{\alpha\beta\gamma}E_{\beta}E^*_{\gamma}$, where $\lambda_{\alpha\beta\gamma}$ is the third-order tensor acquiring non-zero components
in non-centrosymmetric materials. In non-gyrotropic semiconductor materials, the (rectified) photoinduced electric current occurs as a second-order response to a linearly polarized external EM field.
This constitutes the PGE.
This effect does not directly relate to light pressure, the photon-drag phenomenon, or the non-uniformity of the sample or light field intensity, like the photoinduced Dember effect.
Within the D$_{3h}$ point symmetry
group, the third-order conductivity (or transport coefficient) tensor possesses only one nonzero component.
Thus, phenomenologically, the PGE current can be expressed as
\begin{gather}\label{bare1}
j_x=\lambda(|E_x|^2-|E_y|^2),\,\,\,
j_y=-\lambda(E_xE_y^*+E^*_xE_y).
\end{gather}
Therefore, our main task comes down to calculating the coefficient $\lambda$, considering the contribution of electron-hole interaction.


The Boltzmann equations, describing (i) the electron and hole scattering on impurities and (ii) the inter-particle scattering in the framework of the two-band model, read:
\begin{gather}\label{q1}
\frac{\partial f_{\bf p}}{\partial t}-{\bf F}(t)\cdot\frac{\partial f_{\bf p}}{\partial {\bf p}}
=Q_{ei}\{f_{\bf p}\}+Q_{eh}\{f_{\bf p}\},
\\
\frac{\partial f_{\bf k}}{\partial t}+{\bf F}(t)\cdot\frac{\partial f_{\bf k}}{\partial {\bf k}}
=Q_{hi}\{f_{\bf k}\}+Q_{he}\{f_{\bf k}\},
\end{gather}
where ${\bf F}(t)={\bf F}e^{-i\Omega t}+{\bf F}^*e^{i\Omega t}$ is the time-dependent force acting on the mobility electrons and holes of the Dirac monolayer exposed to an external linearly polarized probe EM field $\textbf{E}=(E_x, E_y)$, thus ${\bf F}=e{\bf E}$, $e>0$.
Throughout the paper, the momenta ${\bf k}$ and ${\bf k}'$ refer to holes, and ${\bf p}$ and ${\bf p}'$ to electrons (for clarity and in order to avoid mistakes).
The right-hand sides of these Boltzmann equations contain the collision integrals, describing the particle-impurity and inter-particle interactions.
In general, the r.h.s. of Eq.~\eqref{q1} should also contain e-e and h-h interactions.
However, an analysis shows that in the framework of a two-band model with equal masses of electrons and holes, these contributions vanish~\cite{SMBG}.
Furthermore, let us limit ourselves to considering electron and hole scattering on short-range impurities and assume the corresponding scattering times are identical and independent of the electron and hole energies.
In this case, it is easier to implement the relaxation time approximation for the particle-impurity scattering,
\begin{eqnarray}
\label{Eqq21}
&&Q_{e(h)i}\{f_{{\bf p}({\bf k})}\}=-\frac{f_{{\bf p}({\bf k})}-n_{{\bf p}({\bf k})}}{\tau_i},\\
\label{Eqq22}
&&Q_{eh}\{f_{\bf p}\}=\frac{2\pi}{\hbar S^3}\sum_{{\bf p}',{\bf k}',{\bf k}}|U_{{\bf p}'-{\bf p}}|^2\delta_{{\bf k}'+{\bf p}'-{\bf k}-{\bf p}}\\
\nonumber
&&\times[(1-f_{\bf k})(1-f_{\bf p})f_{{\bf k}'}f_{{\bf p}'}-
(1-f_{{\bf k}'})(1-f_{{\bf p}'})f_{{\bf k}}f_{{\bf p}}]\\
\nonumber
&&~~~~~\times
\delta(\epsilon_{{\bf k}'}+\epsilon_{{\bf p}'}-\epsilon_{{\bf k}}-\epsilon_{{\bf p}}),
\end{eqnarray}
%
where $n_{{\bf p}({\bf k})}$ are equilibrium distribution functions of electrons (holes).
Expression~\eqref{Eqq22} determines the electron-hole collision integral.
Evidently, $Q_{eh}$ is a nonlinear function of the electron and hole distribution functions.
Here, $U_{\bf p}$ is a Fourier transform of the electron-hole interaction potential. In what follows, we will set the sample area $S=1$ in front of the sums for brevity (in most of places). These factors cancel out in the derivations and do not appear in the final formulas.
We also use $\hbar$$=$$k_B$$=$$1$ units and restore dimensionality in the final results.

In Eqs.~\eqref{Eqq21} and~\eqref{Eqq22}, we account for trigonal warping of the electron and hole dispersion: $\epsilon_\mathbf{p}=\epsilon_p^0+w_\mathbf{p}$ with $\epsilon_p^0=p^2/(2m)$, and $w_\mathbf{p}=w_e(p_x^3-3p_xp_y^2)$.
Note, we include the valley index $\eta_e$ into the definition of the warping parameter: $\eta_e w_e\rightarrow w_e$, for brevity.
Thus,
${\bf v}^e({\bf p})={\bf p}/m+w_e\left[3(p_x^2-p_y^2)\mathbf{i}-6p_xp_y\mathbf{j}\right]$
or
${\bf v}^e({\bf p})=(p_x/m+3 w_e(p_x^2-p_y^2),p_y/m-6w_e p_xp_y)$.
The relation for the holes is similar with the replacement $w_e\rightarrow w_h$.

The actual distribution functions entering Eqs.~\eqref{Eqq21} and~\eqref{Eqq22} can be expanded into the powers of external EM field as $f_{\textbf{p}(\textbf{k})}=n_{\textbf{p}(\textbf{k})}+\delta f^{(1)}_{\textbf{p}(\textbf{k})}(t)+\delta f^{(2)}_{\textbf{p}(\textbf{k})}$ with $\delta f^{(1)}_{\textbf{p}(\textbf{k})}(t) = \delta f^{(1)}_{\textbf{p}(\textbf{k})}e^{-i\Omega t}+f^{(1)\ast}_{\textbf{p}(\textbf{k})}e^{i\Omega t}$, thus $\delta f^{(1)}_{\textbf{p}(\textbf{k})}(t)$ and $\delta f^{(2)}_{\textbf{p}(\textbf{k})}$ are the alternating first-order and the stationary second-order corrections to the equilibrium distribution function with respect to external EM field amplitude. The first-order corrections (to electron and hole distribution functions) satisfy the equations:
\begin{eqnarray}
\label{q4}
&&\frac{\partial \delta f^{(1)}_{\bf p}(t)}{\partial t}-{\bf F}(t)\cdot\frac{\partial n_{\bf p}}{\partial {\bf p}}+\frac{\delta f^{(1)}_{\bf p}(t)}{\tau_i}=Q^I_{eh}\{\delta f^{(1)}_{\bf p}(t)\},~~~\\
\label{qq4}
&&\frac{\partial \delta f^{(1)}_{\bf k}(t)}{\partial t}+{\bf F}(t)\cdot\frac{\partial n_{\bf k}}{\partial {\bf k}}+\frac{\delta f^{(1)}_{\bf k}(t)}{\tau_i}=Q^I_{he}\{\delta f^{(1)}_{\bf k}(t)\},~~~
\end{eqnarray}
\begin{widetext}
\begin{eqnarray}
&&Q^I_{eh}\{\delta f^{(1)}_{\bf p}\}=
-2\pi\sum_{{\bf p}',{\bf k}',{\bf k}}
|U_{{\bf p}'-{\bf p}}|^2
\delta(\epsilon_{{\bf k}'}+\epsilon_{{\bf p}'}-\epsilon_{{\bf k}}-\epsilon_{{\bf p}})
\delta_{{\bf k}'+{\bf p}'-{\bf k}-{\bf p}}
\\
\nonumber
&&~~~~~\times\Bigl[\delta f^{(1)}_{\bf p}[(1-n_{\bf k})n_{{\bf k}'}n_{{\bf p}'}+n_{\bf k}(1-n_{{\bf k}'})(1-n_{{\bf p}'})]
-\delta f^{(1)}_{{\bf p}'}[(1-n_{{\bf k}})(1-n_{{\bf p}})n_{{\bf k}'}+n_{\bf k}n_{\bf p}(1-n_{{\bf k}'})]\\
\nonumber
&&~~~~~~~+\delta f^{(1)}_{\bf k}[(1-n_{\bf p})n_{{\bf k}'}n_{{\bf p}'}+n_{\bf p}(1-n_{{\bf k}'})(1-n_{{\bf p}'})]
-\delta f^{(1)}_{{\bf k}'}[(1-n_{{\bf k}})(1-n_{{\bf p}})n_{{\bf p}'}+n_{\bf k}n_{\bf p}(1-n_{{\bf p}'})]\Bigr],
\end{eqnarray}
\end{widetext}
where $Q^I_{eh}\{\delta f^{(1)}_{\bf p}\}$ is a linearized collision integral with respect to the first-order correction to the electron distribution function; $Q^I_{he}\{\delta f^{(1)}_{\bf k}\}$ can be found analogously (see Supplemental Material for details~\cite{[{See the Supplemental Material at http://link.aps.org/supplemental/10.1103/PhysRevB.109.245414 which gives the details of the derivations and refers to~\cite{Zaitsev:2014aa, PhysRevLett.123.095301, Durnev:2018}}]SMBG}).

The stationary vPGE current density is determined by the stationary part of the second-order correction to the particle distribution functions, which satisfies the equations:
\begin{widetext}
\begin{eqnarray}\label{IIq1}
&&-{\bf F}\cdot\frac{\partial \delta f^{(1)*}_{\bf p}}{\partial {\bf p}}-
{\bf F}^*\cdot\frac{\partial \delta f^{(1)}_{\bf p}}{\partial {\bf p}}
+\frac{\delta f^{(2)}_{\bf p}}{\tau_i}
=Q^I_{eh}\{\delta f^{(2)}_{\bf p}\}+
Q^{II}_{eh}\{\delta f^{(1)}_{\bf p}\},\\
\label{IIqq1}
&&{\bf F}\cdot\frac{\partial \delta f^{(1)*}_{\bf k}}{\partial {\bf k}}+
{\bf F}^*\cdot\frac{\partial \delta f^{(1)}_{\bf k}}{\partial {\bf k}}
+\frac{\delta f^{(2)}_{\bf k}}{\tau_i}
=Q^I_{eh}\{\delta f^{(2)}_{\bf k}\}+
Q^{II}_{eh}\{\delta f^{(1)}_{\bf k}\},
\end{eqnarray}
\begin{eqnarray}
&&Q^I_{eh}\{\delta f^{(2)}_{\bf p}\}=-2\pi\sum_{{\bf p}',{\bf k}',{\bf k}}|U_{{\bf p}'-{\bf p}}|^2
\delta(\epsilon_{{\bf k}'}+\epsilon_{{\bf p}'}-\epsilon_{{\bf k}}-\epsilon_{{\bf p}})
\delta_{{\bf k}'+{\bf p}'-{\bf k}-{\bf p}}\\
\nonumber
&&~~~~~~~~~~~~\times\Bigl[\delta f^{(2)}_{\bf p}[(1-n_{\bf k})n_{{\bf k}'}n_{{\bf p}'}+n_{\bf k}(1-n_{{\bf k}'})(1-n_{{\bf p}'})]
-\delta f^{(2)}_{{\bf p}'}[(1-n_{{\bf k}})(1-n_{{\bf p}})n_{{\bf k}'}+n_{\bf k}n_{\bf p}(1-n_{{\bf k}'})]\\
\nonumber
&&~~~~~~~~~~~~~~+\delta f^{(2)}_{\bf k}[(1-n_{\bf p})n_{{\bf k}'}n_{{\bf p}'}+n_{\bf p}(1-n_{{\bf k}'})(1-n_{{\bf p}'})]
-\delta f^{(2)}_{{\bf k}'}[(1-n_{{\bf k}})(1-n_{{\bf p}})n_{{\bf p}'}+n_{\bf k}n_{\bf p}(1-n_{{\bf p}'})]\Bigr],
\\
&&Q^{II}_{eh}\{\delta f^{(1)}_{\bf p}\}=-2\pi\sum_{{\bf p}',{\bf k}',{\bf k}}|U_{{\bf p}'-{\bf p}}|^2
\delta(\epsilon_{{\bf k}'}+\epsilon_{{\bf p}'}-\epsilon_{{\bf k}}-\epsilon_{{\bf p}})
\delta_{{\bf k}'+{\bf p}'-{\bf k}-{\bf p}}
\\
\nonumber
&&~~~~~~~~~~~~~~~\times\Bigl[\overline{\delta f^{(1)}_{\bf p}(t)\delta f^{(1)}_{\bf k}(t)}
[(1-n_{{\bf k}'})(1-n_{{\bf p}'})-n_{{\bf k}'}n_{{\bf p}'}]
-\overline{\delta f^{(1)}_{{\bf p}'}(t)\delta f^{(1)}_{{\bf k}'}(t)}
[(1-n_{{\bf k}})(1-n_{{\bf p}})-n_{{\bf k}}n_{{\bf p}}]\Bigr],
\end{eqnarray}
\end{widetext}
where $\overline{A(t)B(t)}$ stands for the time averaging; $Q^{II}$ combines the second-order terms after the linearization of the e-h collision integral.
Solving Eqs.~\eqref{q4}-\eqref{IIqq1} by successive approximations with respect to the particle collision integrals, we come up with the Coulomb-induced corrections as
$\delta f^{(2)}_{\bf p}=\delta_0 f^{(2)}_{\bf p}+\delta_C f^{(2)}_{\bf p}$ and
$\delta f^{(2)}_{\bf k}=\delta_0 f^{(2)}_{\bf k}+\delta_C f^{(2)}_{\bf k}$,
where
\begin{gather}\label{IIq2a}
\delta_0 f^{(2)}_{\bf p}=\tau_i\left({\bf F}\cdot\partial_{\bf p} \delta_0 f^{(1)*}_{\bf p}+
{\bf F}^*\cdot\partial_{\bf p} \delta_0 f^{(1)}_{\bf p}\right),\\
\label{IIq2b}
\delta_0 f^{(2)}_{\bf k}=-\tau_i\left({\bf F}\cdot\partial_{\bf k} \delta_0 f^{(1)*}_{\bf k}+
{\bf F}^*\cdot\partial_{\bf k} \delta_0 f^{(1)}_{\bf k}\right),
\end{gather}
and thus,
\begin{eqnarray}
\label{IIq3}
&&\delta_C f^{(2)}_{\bf p}=\tau_i\left({\bf F}\cdot\partial_{\bf p} \delta_C f^{(1)*}_{\bf p}+
{\bf F}^*\cdot\partial_{\bf p} \delta_C f^{(1)}_{\bf p}\right)\\
\nonumber
&&~~~~~~~~~+\tau_i\left(Q^I_{eh}\{\delta_0 f^{(2)}_{\bf p}\}+
Q^{II}_{eh}\{\delta_0 f^{(1)}_{\bf p}\}\right),\\
\label{IIq32}
&&\delta_C f^{(2)}_{\bf k}=-\tau_i\left({\bf F}\cdot\partial_{\bf k} \delta_C f^{(1)*}_{\bf k}+
{\bf F}^*\cdot\partial_{\bf p} \delta_C f^{(1)}_{\bf k}\right)
\\
\nonumber
&&~~~~~~~~~+\tau_i\left(Q^I_{he}\{\delta_0 f^{(2)}_{\bf k}\}+
Q^{II}_{he}\{\delta_0 f^{(1)}_{\bf k}\}\right).
\end{eqnarray}
%
%
Expressions~\eqref{IIq3} and~\eqref{IIq32} describe full e-h scattering corrections to the electric current density,
\begin{eqnarray}
j_\alpha=
e
\sum_{\bf k}v^h_\alpha({\bf k})\delta_C f^{(2)}_{\bf k}
-
e
\sum_{\bf p}v^e_\alpha({\bf p})\delta_C f^{(2)}_{\bf p}.
\end{eqnarray}
%
Without the loss of generality, we can choose a particular direction of the force to make our derivations clear: $\mathbf{F}=(F_x,0)$, and then,
\begin{eqnarray}
j_x=
e
\sum_{\bf k}v^h_x({\bf k})\delta f^{(2)}_{\bf k}
-
e
\sum_{\bf p}v^e_x({\bf p})\delta f^{(2)}_{\bf p}.
\end{eqnarray}

It is more convenient to split contributions~\eqref{IIq3} and~\eqref{IIq32} into several terms and consider them separately (see Supplemental Material~\cite{SMBG} for details).
Performing the calculations, we find:
\begin{eqnarray}
\label{EqMainGeneral}
&&j_x=
(w_e-w_h)
\frac{141}{32}
\frac{e}{\hbar}
\frac{\tau_i^3F_x^2}{1+\Omega^2\tau_i^2}
\frac{n_en_h}{\sqrt{mT^3}}
\\\nonumber
&&~~~~~~~~~~~~~\times
\frac{1}{\sqrt{\pi}}
\int\limits_0^\infty dq
|U_q|^2
q^2e^{-\frac{q^2}{4mT}},
\end{eqnarray}
where $n_e$ and $n_h$ are electron and hole equilibrium densities at a given temperature (we restored the Planck's and other constants).
Formula~\eqref{EqMainGeneral} is a general expression for the Coulomb interaction-induced vPGE current.
When applied to a TMD MoS$_2$ monolayer, $U_q$ takes the form of the Keldysh-Rytova potential~\cite{Rytova,Keldysh-R}:
$U_{q}=2\pi e^2/\varepsilon q(1+2\pi\alpha_\textrm{2D}q)$, where $\alpha_\textrm{2D}$ is a polarizability of the monolayer, and $\varepsilon$ is dielectric constant of surrounding media. This form of particle-particle interaction is general, and at $\alpha_{2D}=0$, it turns into the bare Coulomb interaction.
Introducing the dimensionless variable $x=q/\sqrt{4mT}$, we find electron-hole interaction-induced contribution to $\lambda$, describing vPGE current density:
\begin{eqnarray}
\label{EqMain}
&&\lambda=(w_e-w_h)
\frac{141}{16}
\frac{e^3}{\hbar}
\frac{\tau_i^3}{1+\Omega^2\tau_i^2}
\frac{n_en_h}{T}
\left(\frac{2\pi e^2}{\varepsilon}\right)^2
{\cal J},
\\
\nonumber
&&{\cal J}=\frac{1}{\sqrt{\pi}}
\int\limits_0^\infty
dx
\left[
\frac{1}{1+x\,2\pi\alpha_\textrm{2D}\sqrt{4mT}}\right]^2
e^{-x^2}.
\end{eqnarray}
%
Formula~\eqref{EqMain} is the central result of this paper.
It should be mentioned that the dimensionless integral ${\cal J}$ can be estimated numerically or even found analytically, but the resulting expression is rather complicated, and thus, we do not present it here.


\section{Results and discussion}

Formula~\eqref{EqMain} potentially has two important limiting cases due to the presence of two effective distances: $\alpha_\textrm{2D}$ and $\hbar/\sqrt{4mT}$ (in the power of the exponent).
The estimation shows that {$\alpha_\textrm{2D}\sqrt{4mT}\ll 1$ up to the Debye temperatures (and even higher), when particle-phonon scattering starts to play the dominating role.
Thus, we can safely disregard the term proportional to $\alpha_\textrm{2D}$.
Then, the integral over $x$ turns into the Euler-Poisson integral, which gives
\begin{eqnarray}
\label{EqMainSimplified}
\lambda=(w_e-w_h)
\frac{141}{32}
\frac{e^3}{\hbar}
\frac{\tau_i^3}{1+\Omega^2\tau_i^2}
\left(\frac{2\pi e^2}{\varepsilon}\right)^2
\frac{n_en_h}{T}.
\end{eqnarray}
Let us mention that in~\eqref{EqMainSimplified}, all the temperature dependence is in the fraction containing the concentrations $n_e$, $n_h$, and $T$.

Furthermore, let us compare Eq.~\eqref{EqMainSimplified}  with the bare vPGE effect, reported elsewhere~\cite{Entin2019}.
Combining together the bare vPGE current and the Coulomb-induced correction to the current yields
\begin{eqnarray}
\label{EqBarePGE}
&&\tilde\lambda=12\frac{e^3\tau_i^2}{1+\Omega^2\tau_i^2}
\left[(n_hw_h-n_ew_e)\right.
\\
\nonumber
&&~~~~\left.-(w_h-w_e)C_0\frac{(2\pi)^2}{\hbar}\frac{e^4n_en_h}{(2\varepsilon T)^2}(T\tau_i)\right],
\end{eqnarray}
where $C_0\approx0.37$, and we use $\tilde\lambda$ instead of $\lambda$ to distinguish between the Coulomb-related and full (bare plus Coulomb force-induced) vPGE currents.
Evidently, the dependence on the particle densities of the first and the second terms here differ.
The dependence on the product $n_en_h$ of the second term is an attribute of the Coulomb-related contribution.

In order to clearly understand the physics of the Coulomb-induced term, let us take into account that the equilibrium densities in an intrinsic Dirac material equal $n_e=n_h=n_i$, where $n_i=(mT/2\pi\hbar^2)\exp(-\Delta/2T)$ is an intrinsic particle density with $\Delta$ the material bandgap.
Thus, the square brackets in Eq.~\eqref{EqBarePGE} read as
\begin{eqnarray}
\label{SomEqnee}
n_i(w_h-w_e)\left[1-C_0\frac{(2\pi)^2}{\hbar}\left(\frac{e^2\sqrt{n_i}}{2\varepsilon T}\right)^2(T\tau_i)\right].
\end{eqnarray}
We immediately conclude that the second term is opposite to the bare vPGE contribution~\cite{[{It should be noted, that the expression for bare vPGE contains the square of the momentum relaxation time (see the first term in Eq.~\eqref{EqBarePGE}), while the Coulomb-induced correction Eq.~\eqref{EqMainSimplified}, at first glance, behaves as $\tau_i^3$.
This is a seeming contradiction since the particle-particle scattering time $\tau^{-1}_{eh}\sim e^4n_i/T$, thus the Coulomb interaction-induced term yields $\tau_i^3/\tau_{eh}\sim\tau^2$ scaling behaviour}]FTN2}.
It results in suppression of the vPGE due to Coulomb interaction between electrons and holes.
Such behavior can be qualitatively explained by the following argument: electrons and holes experience friction (or a coulomb drag) since they move in opposite directions.
To qualitatively estimate the Coulomb contribution, an estimation of intensity is necessary.

In the second term in square brackets of Eq.~\eqref{SomEqnee}, the factor $e^2\sqrt{n_i}/2T\ll 1$ is the relation of the Coulomb interaction energy of the electron and the hole to their kinetic energy given by temperature for non-degenerate gas statistics.
Such behavior is expected because we considered the Coulomb-related contribution to vPGE perturbatively.
However, the factor $T\tau_i/\hbar$ in Eq.~\eqref{SomEqnee} reflects the relation between particles' kinetic energy and their broadening due to scattering on impurities.
The latter should be weak compared to the particle energy, thus $T\tau_i/\hbar\gg 1$.
This factor compensates for the smallness of the preceding factor and can even result in an essential suppression of the vPGE due to electron-hole scattering.

Furthermore, the general result~\eqref{EqBarePGE} only holds in the equilibrium.
The net vPGE current should be summed over two non-equivalent valleys, and it vanishes due to the time-reversal symmetry.
Indeed, restoring the valleys indices $\eta_e=\pm1$ and $\eta_h=\mp1$ for electron and hole warping amplitudes, $w_e\rightarrow \eta_ew_e$, $\,w_h\rightarrow \eta_hw_e$, and summing up over $\eta_e=\pm1$ and $\eta_h=\mp1$ at equilibrium densities $n_e^{\eta_e}=n_h^{\eta_h}=n_i$ in both valleys, yields zero net vPGE current.
In order to achieve a nonzero net current, the time-reversal symmetry is to be destroyed.
It can be done by sample illumination with circularly-polarized pump EM field with frequency producing interband transitions populating a given valley (at given circular polarization), see, e.g.,~\cite{PhysRevB.77.235406}, and leaving the other valley with equilibrium populations, $n_i$.

Let the valley with $\eta_e=-\eta_h=1$ be populated with densities $n_e=n_h=n_i+\delta n_{ph}$, where $\delta n_{ph}$ is the photoinduced density correction.
If electrons and holes in the other valley   $\eta_e=-\eta_h=-1$ have equilibrium densities $n_i$, the net vPGE current is determined by the coefficient $\tilde{\lambda}$ in the form:
\begin{eqnarray}
\label{netPGE}
&&\tilde{\lambda}=-12\frac{e^3\tau_i^2}{1+\Omega^2\tau_i^2}(w_h+w_e)
\\
\nonumber
&&~~~~~~\times\left[1-C_0\frac{(2\pi)^2}{\hbar}\frac{e^4(2n_i+\delta n_{ph})}{(2\varepsilon T)^2}(T\tau_i)\right]\delta n_{ph}.
\end{eqnarray}
At low intensity of photogeneration, $n_i\gg\delta n_{ph}$, the total current increases linearly with the circular light intensity, $\delta n_{ph}\sim I$.
With increasing the intensity when the density of non-equilibrium carries exceeds the density of equilibrium ones, $n_i\ll\delta n_{ph}$, the Coulomb-induced term increases as $I^2$, and it can totally suppress the vPGE effect.

The developed formalism is applicable to two types of systems.
(i) It can be directly applied to small-bandgap materials, such as gapped graphene, where the intrinsic charge density can be rather large, and two cases, $n_i\gg\delta n_{ph}$ and $n_i\ll\delta n_{ph}$, are possible.
(ii) In the case of intrinsic large-bandgap materials, such as TMDs, the intrinsic density at reasonable temperatures is exponentially small, and vPGE is determined exclusively by the density of optically generated carriers of charge, $\delta n_{ph}$, in one of the valleys.

Let us estimate the Coulomb-induced vPGE for small-gap materials considering the gapped graphene as a testing ground.
Assuming $n_i\gg\delta n_{ph}$ and taking $n_i=5\cdot10^7$~cm$^{-2}$, $T=30$~K, $\tau_i=10^{-12}$~s, $\varepsilon=3$, the second term in the square brackets in Eq.~\eqref{netPGE} is $\approx 0.26$.
Then, due to the factor $T\tau_i/\hbar>1$, the Coulomb-induced term may sufficiently suppress the non-interacting particle contribution.

Furthermore, let us discuss the applicability of the two-band model.
In the case of gapped graphene, the gap is induced when the monolayer is placed on a substrate.
In this case, two bands naturally emerge and the two-band model is perfectly justified to study the effect of electron-hole scattering processes on the conductivity of intrinsic graphene~\cite{PhysRevB.109.085424}.
In the case of a TMD monolayer, its band structure consists of the spin-orbit split subbands both in the conduction and valence bands.
The intrinsic electron-hole densities here are exponentially small due to large bandgap. Hence, the vPGE may only exist in the presence of photoinduced electrons and holes. The interband optical transitions conserve the spin, and thus, the photogenerated carriers effectively occupy only one spin subband in conduction and one spin subband in the valence band.
In the case of large spin relaxation times, the two-band model is also an adequate approximation.





\section*{Conclusions}
We developed a theory of the Coulomb interaction-related contribution to the valley photogalvanic effect in two-dimensional non-centrosymmetric Dirac materials possessing the nontrivial structure of valleys and exposed to an external electromagnetic field, taking gapped graphene and MoS$_2$ as examples.
The valley photogalvanic effect is the result of two factors: (i) the presence of the trigonal warping of electrons and holes' bands and (ii) the Coulomb interaction between the carriers of charge.
We considered the low-frequency limit of the external linearly-polarized probe field: the field frequency is smaller than the temperature $T$, and the electron-electron and electron-hole scattering times are much larger than the particle-impurity scattering time.
Using the Boltzmann transport equations, we demonstrated that the electron-hole scattering dominates electron-electron scattering in intrinsic semiconductors and it might be comparable with the bare valley photogalvanic effect.
We found, that the Coulomb interaction-related contribution to the valley photogalvanic current is opposite to the bare valley photogalvanic current.

\vspace*{2mm}

\acknowledgements

\vspace*{1mm}

We were supported by the Institute for Basic Science in Korea (Project No.~IBS-R024-D1), Ministry of Science and Higher
Education of the Russian Federation (Project FSUN-2023-0006), and the Foundation for the Advancement of Theoretical Physics and Mathematics ``BASIS''.





\bibliography{biblio}
\bibliographystyle{apsrev4-2}

\end{document}